\documentclass[conference]{IEEEtran}
\IEEEoverridecommandlockouts

\usepackage{cite}
\usepackage{amsmath,amssymb,amsfonts}
\usepackage{algorithmic}
\usepackage{graphicx}
\usepackage{textcomp}
\usepackage{xcolor}
\def\BibTeX{{\rm B\kern-.05em{\sc i\kern-.025em b}\kern-.08em
    T\kern-.1667em\lower.7ex\hbox{E}\kern-.125emX}}

\usepackage[english=nohyphenation]{hyphsubst}

\usepackage{makecell}

\newcommand{\rev}[1]{{\color{black} #1}}
\newcommand{\revv}[1]{{\color{black} #1}} 

\graphicspath{{./figs/}}

\renewcommand\IEEEkeywordsname{Keywords}

\begin{document}

\title{
A labeled dataset of cloud types using data from GOES-16 and CloudSat\\



}

\author{\IEEEauthorblockN{Paula Romero Jure}
\IEEEauthorblockA{\textit{Gerencia de Vinculaci\'on Tecnol\'ogica} \\
\textit{CONAE} \\
Cordoba, Argentina \\
paula.romero@mi.unc.edu.ar}
\and
\IEEEauthorblockN{Sergio Masuelli}
\IEEEauthorblockA{\textit{Grupo de Física de la Atmósfera} \\
\textit{FaMAF - UNC}\\
C\'ordoba, Argentina \\
smasuelli@unc.edu.ar}
\and
\IEEEauthorblockN{Juan Cabral}
\IEEEauthorblockA{\textit{Gerencia de Vinculaci\'on Tecnol\'ogica} \\
\textit{CONAE}\\
C\'ordoba, Argentina \\
jbcabral@unc.edu.ar}
}

\maketitle

\IEEEpubidadjcol

\begin{abstract}

In this paper we present the development of a dataset consisting of 91 \revv{Multi-band Cloud and Moisture Product Full-Disk (MCMIPF)} from the \revv{Advanced Baseline Imager (ABI) on board} GOES-16 geostationary satellite with 91 temporally and spatially corresponding CLDCLASS products from the CloudSat polar satellite. The products are diurnal, corresponding to the months of January and February 2019 and were chosen such that the products from both satellites can be co-located over \revv{South America}. The CLDCLASS product provides the cloud type observed for each of the orbit's steps and the GOES-16 multiband images contain pixels that can be co-located with these data. We develop an algorithm that returns a product in the form of a table that provides pixels from multiband images labeled with the type of cloud observed in them. These labeled data conformed in this particular structure are very useful to perform supervised learning. This was corroborated by training a simple linear artificial neural network based on the work of \rev{Gorooh} et al. \revv{(2020)}, which gave good results especially for the classification of deep convective clouds \cite{Gorooh2020}.

\end{abstract}


\section{Introduction}


Clouds are a vital component of Earth's atmosphere as they make a significant contribution in the regulation of water cycles and the modulation of radiative fluxes. The World Meteorology Organization (WMO) defines clouds as hydrometeors consisting of particles of water and ice suspended in the atmosphere and usually not touching the ground. According to the cloud's appearance, composition, altitude and other features, they can be classified into ten main groups or \textit{genera} which are named in Latin after a ``main characteristic" \cite{wmo}.

Obtaining information about the distribution of clouds and their types is critical not only for weather forecasting and climate monitoring. It can also make a significant contribution in the improvement of precision in climate models, given that ``clouds and aerosols continue to contribute the largest uncertainty to estimates and interpretations of the Earth’s changing energy budget'' \cite{intergovernmental}. 

In order to accomplish this task, analyzing data retrieved from  remote sensors located on board artificial satellites orbiting the Earth is a widely used method. There are different kinds of sensors as well as different kinds of orbits, both are planned together depending on the scientific goal of the mission. 

\revv{Monostatic radars are active sensors that produce a signal on} a fixed frequency that is transmitted through the atmosphere by an antenna and after it interacted with an object, it's detected by the same antenna \cite{matzel}. The Cloud Profiling Radar (CPR) is an example of such instruments, \revv{which operates on a frequency of $94 GHz$.} It is located \revv{on CloudSat} \cite{stephens2002cloudsat}, an artificial satellite with a polar sun-synchronous orbit with a characteristic period of 99 minutes. 

\revv{Radiometers are passive sensors that are able to detect radiation} from the Earth in a determined range of wavelengths. For instance, the  Advanced Baseline Imager (ABI) is a multi-spectral radiometer on board the ``Geostationary Operational Environmental Satellite'' (GOES-16, \cite{galica2016goes}). Given that its orbit is geostationary, with a central longitude of \rev{$-75°$} and an altitude of 35786 km, the sensor is able to \revv{acquire} images of nearly the whole American continent, in particular of South America.

For our study, the information from CloudSat and GOES-16 are complementary. As CPR's signal can penetrate clouds, it can \revv{provide} information of the vertical profile with the possibility to retrieve information, along \rev{its} track, about the phase of the hydrometers \rev{allowing it} to classify clouds. \revv{However, it does not acquires data in the cross-track direction, i.e. rather than an image, the instrument generates a transect.} On the other hand, ABI measures radiance and then images are generated from them, but for its spectral bands (visible to long wave infrared) clouds are in general opaque, in consequence it can sense only the highest layers of the clouds.

Therefore, data about cloud types in the highest layers obtained by CPR can be used as tags for pixels of images obtained by ABI. This process, in which we are comparing co-located measurements taken by two different satellite's instruments, is widely used for intercalibration purposes.

Usually, geostationary satellites are excellent candidates to be collocated with polar-orbit satellites because the formers remain always in a static position relative to some point of the Earth while the latters will pass through the area that the geostationary satellites observe some times per day \cite{chander}. By this means, GOES-16 and CloudSat data can be collocated.

The main objective of this work was to develop a dataset in which pixels of a GOES-16 image are labelled with the cloud types that can be observed in CloudSat.

This paper is organized as follows: In Section~\ref{section:bg} we present other works in which datasets related to clouds properties were developed, some of them using CloudSat products, in Section~\ref{section:matandmeth}, we describe the products of CloudSat and GOES-16 that we used throughout the work, and the algorithm developed to obtain the co-located ``final dataset''. We describe the obtained dataset in Section~\ref{section:results}, with an small example \rev{on} how to use it in a automatic classifier in order to corroborate the quality of the data. Finally in  Section~\ref{section:conclu} we provide the conclusions an the future work.

\section{Background}
\label{section:bg}

Clouds are distinguished from each other by their vertical extent, particle size and phase. 
As radiative transfer theory states, when electromagnetic radiation interacts with them, the differences between different types of clouds produce different results depending on the wavelength of the interacting radiation \cite{prap}. 
By observing the clouds in different spectral bands, suitably chosen, algorithms can be designed to detect the differences sensitive to different spectral lengths \cite{kidder, marshak}.

CPR provides a cloud classification product, 2B-CLDCLASS, which has been used in combination with other sensors to improve cloud studies. Specifically, \rev{Gorooh} et al. \cite{Gorooh2020}, used this product as a validation to train an unsupervised cloud classification neural network for MODIS images. Zantedeschi et al. \revv{(2019)} used this product combined with MODIS products to create CUMULO, a dataset consisting of one year of 1km resolution MODIS multi-spectral imagery merged with pixel-width ‘tracks’ of CloudSat cloud labels \cite{cumulo}.

Other examples of data sets about clouds and their properties include the ``CLoud property dAtAset using SEVIRI'' \cite{claas1} (CLAAS), which is composed of cloud products at different processing levels from pixel-based data to daily and monthly summaries. The dataset was created using data from Europe collected by SEVIRI, a 12-channel passive imager on board the MSG geostationary satellites operated by EUMETSAT, in a time slot of 8 years, also an improvement of the data set, CLAAS-2, was released in 2017\cite{claas2}. 

We see that all of these data sets use products from at least two satellites. Correlate data from two different satellite instruments is a widely used process. It is used, for example, to perform instrument intercalibrations, which is an effective and widely used method to verify the coincidence of two measurements of the same space and at the same time made by two different instruments \cite{chander}. \revv{Collocations between two satellite sensors are occasions where both sensors observe the same place at roughly the same time \cite{collocations}. Although this is an ideal situation and the sensors are never exactly aligned, thresholds are usually applied to define the collocations, which impacts in the uncertainty of the comparison, partially due to the scene variability within the range of the collocation criteria \cite{chander}.}


\section{Materials and Methods}
\label{section:matandmeth}

To perform this work we need different observations of the same area, these data are obtained from ABI images. On the other hand it is also necessary to obtain a classification of the clouds in the same observation area, this is provided by the CloudSat satellite. In this section, we will explain the algorithm developed to obtain a table of image pixels co-located with the type of cloud seen in them. This forms what for this work is considered the ``final product''.

\subsection{CloudSat data}

As mentioned earlier, CPR is a monostatic radar with a frequency of $94 GHz$ on board CloudSat, that \revv{acquires} vertical profiles of the atmosphere, in particular of clouds, with a nominal resolution of $500 km$ and  in a footprint of $1.4 \times 2.5 km$ (cross and along track) \cite{cloudsat1}. \revv{CloudSat has been in daytime-only operations since 2011 due to battery malfunction, requiring sunlight to power the radar \cite{cloudsat1}. Therefore, only daytime data have been useful since that year.}

In this work we use products called 2B-CLDCLASS which are level 2 processing products, which is delivered using Hierarchical Data Format (HDF) \cite{folk2011overview}. 2B-CLDCLASS provides two types of data: the so-called ``Data Fields'' containing meta-information about the observation, and the observation itself called ``Geolocation Fields''. From the first ones we are interested in \textit{CloudLayerType} which provides the types of clouds spotted at different layers of the atmosphere, and from the second ones \textit{Latitude}, \textit{Longitude} and \textit{UTC start} (observation time of each latitude and longitude).

We only use the highest layer, since it is the one that can be measured by the ABI. The possible cloud types, corresponding to the eight WMO classes plus a ninth class indicating the absence of cloudiness, are shown in Table~\ref{tab:tiposnubes}.
    
    \begin{table}[b]
    \caption{Classes distinguished by product 2B-CLDCLASS.}
    \begin{center}
    \begin{tabular}{llr}
    \hline
    \thead{\textbf{Class}} & \thead{\textbf{Height in atm} \cite{iribarne:chcloud}} & \thead{\textbf{ID}}  \\
    \hline \\
    No clouds (No) & -- & 0  \\ 
    Cirrus (Ci) & High  & 1 \\
    Altocumulus (Ac) & Medium  & 2 \\
    Altostratus (As) & Medium  & 3 \\
    Nimbostratus (Ns) & Low  & 4 \\
    Stratocumulus (Sc) & Low  & 5 \\
    Stratus (St) & Low  & 6 \\
    Cumulus (Cu) & Convective  & 7 \\
    Deep Convection (DC) & Convective  & 8 \\
    \hline
    \end{tabular}
    \label{tab:tiposnubes}
    \end{center}
    \end{table}

\subsection{GOES-16 data}

\begin{figure}[b].
    \centering
    \includegraphics[width=6cm]{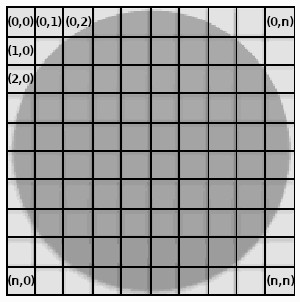}
    \caption{Schematic illustration of an image of an ABI band. It can be represented as a 2D square grid of n pixels high by n pixels wide where each square is a pixel and can be located by the row and column of the grid in which it is located. The size of the image is given by the spatial resolution of the band but for the MCMIPF product, n=5424 for all bands.}
    \label{fig:goes grid}
\end{figure}

The ABI is a passive imager onboard GOES-16. It images in 16 bands. Bands 1 to 6 are reflective, bands 8 to 16 are emissive and band 7 is a medium band since its central wavelength (WL) is $3.8 \mu m$. All bands have a temporal resolution of $5$, $10$ or $15$ minutes depending on the availability of the radiometer modes. Each band has a defined Spatial Resolution (SR), which can be seen in Table~\ref{tab:goes}. Bands with a spatial resolution of $2 km$ produces images of size $5424 \times 5424$. A schematic illustration of how an ABI image looks, for one band, is shown in Fig. \ref{fig:goes grid}.

Level 2 processing products with Cloud and Moisture (CMI) calls were used to create the dataset. In this product, 
\revv{\begin{quote}
    the reflectance measured by ABI in the visible bands is converted to brightness values and the radiance measured in infrared bands is converted to brightness temperatures \cite{pug5}.
\end{quote} }
\revv{CMI files are generated for each of the sixteen ABI reflective and emissive bands, each one with its corresponding spatial resolution (SR) as listed in Table 1. The GOES-16 team also provides a multiband CMI file, calles ''Cloud and Moisture Imagery Product Full-Disk´´ (MCMIPF) file for the Full-Disk mode. This product includes CMI images from all sixteen ABI bands, all unified at the same spatial resolution of $2 km$. These CMI products are delivered using the netCDF-4 file format \cite{rew2006netcdf}.
}

\begin{table}[t]
    \caption{ABI bands, their main use relevant to this work (what they support the characterization of), their central wavelength ($\mu m$) and their spatial resolution (km).}
    \begin{center}
    \begin{tabular}{lrrr}
    \hline
    \thead{\textbf{Usage}}&\thead{\textbf{Band}}&\thead{\textbf{WL($\mu m$)}}&\thead{\textbf{SR(km)l}}\\
    \hline  \\
    Aerosols &1& 0.47 & 1 \\
    Daytime clouds, fog. &2&0.64 & 0.5 \\
    High water/land contrast.&3& 0.87 & 1\\
    Daytime cirrus cloud.&4& 1.38 & 2\\
    Daytime cloud-top phase and part. size.&5& 1.61 & 1 \\
    Daytime land, cloud properties, part. size.&6& 2.25 & 2\\
    Surface and cloud. &7& 3.89 & 2\\
    High-level atmospheric WV, rainfall.&8& 6.17 & 2\\
    Mid-level atmospheric WV, rainfall.&9& 6.93 & 2\\
    Lower-level WV.  &10& 7.34 & 2\\
    Cloud phase, dust, rainfall.&11& 8.44 & 2\\
    Total ozone, turbulence.&12& 9.61 & 2\\
    Surface and clouds.&13& 10.33 & 2\\
    Clouds, rainfall.&14& 11.19 & 2\\
    Moisture, clouds.&15& 12.27 & 2\\
    Air temperature, clouds.&16& 13.27 & 2\\
    \hline
    \end{tabular}
    \label{tab:goes}
    \end{center}
    \end{table}

\subsection{Time selection and co-location}
    
\revv{The need then arises to unify the same positions on two satellites, one geostationary and one polar, in the same coordinate system. In the case of CloudSat, the time, latitude and longitude of each step of the track are provided. In ABI images the location of each pixel is given in a stationary projection, as the GOES-16 satellite is geostationary, and can be converted to a projection in latitude and longitude coordinates as provided in CloudSat products.} In this context we implemented an algorithm that computes this coordinate change based on the GOES-16 user manual \cite{pug2}.

Ninety-two CloudSat 2B-CLDCLASS products with corresponding GOES-16 MCMIPF products from among those available during the day for January and February 2019 were used to compile the dataset. The year of study was arbitrarily selected among which there is an intersection of operational data from both satellites: MCMIPF products are available since October 2017, while CloudSat operational data are available since 2006. As of today, 2B-CLDCLASS products are available until August 2020, while MCMIPF products are available \revv{in real time} after being taken by ABI, which is operational 24 hours a day. On the other hand, the months of study were selected because in South America, the area chosen for the analysis, it is the time of the year when there is more rain and storms, so there is a greater variability in the types of clouds.
    
Only daytime products were used, a schedule conditioned by the CloudSat satellite due to its battery failure. The time slot when CloudSat passes over South America during the day is between 15:00 and 18:30. Given all these conditions, there are between one and two pairs of CLDCLASS-MCMIPF products per day available.

The 91 product pairs were downloaded, inspected and re-uploaded as a dataset to the machine learning social network Kaggle \cite{kaggle} to make it more accessible. The size of the dataset is $40 GB$. 
    
\subsection{Algorithm}
    
To locate each step of the CloudSat orbit on a GOES-16 image, we developed a coordinate change algorithm. Each pixel of a GOES-16 image can be identified with a coordinate (row, column) since the image is like a grid of size $n \times n$ as shown in Fig~\ref{fig:goes grid}. Each coordinate corresponds to a fixed location on Earth, which can be geolocated in a geostationary \revv{projection}. Then, it is possible to perform a coordinate change between this system and the one provided by the 2B-CLDCLASS product where each step of the CloudSat orbit is identified with a coordinate (Latitude, Longitude). The algorithm developed and validated in the GOES-16 user's manual \cite{pug2}.

\revv{Then, we transform the (Lat, Lon) coordinates of each step of the CloudSat orbit into (row,col) coordinates of the ABI images, where row is each row of pixels in an image and col is each column of pixels:} \\ $(Lat, Lon) \xrightarrow{} (x_{geos}, y_{geos})  \xrightarrow{} (row,col)$.
    
\begin{figure}[b]
    \centerline{\includegraphics[width=\linewidth]{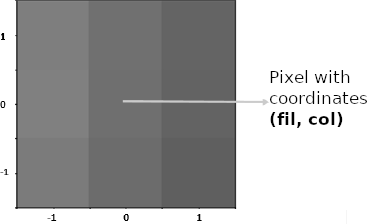}}
    \caption{Example of a cropped image. This crop was extracted from a band 13 image and it is normalized according to \ref{eq:norm}. The central pixel has location (Row,Col) in the entire image and the choice of this location was made knowing that the CloudSat satellite flew over it at the same time (except for a threshold) that the ABI radiometer was taking the image. }
    \label{fig:figure_ref}
\end{figure}

Once the (Column, Row) point was obtained, a MCMIPF product cutout of size 3x3 was extracted for each of the 16 bands, in which the central pixel is the (Column, Row) point. An example is shown in figure \ref{fig:figure_ref}. Then, this cropped image was normalized according to \ref{eq:norm}.
    
\begin{equation}\label{eq:norm}
    \textbf{x}_{norm}(\textbf{x}^j)= \frac{\textbf{x}^j - min(\textbf{C}^j)}{max(\textbf{C}^j)- min(\textbf{C}^j)}, 
\end{equation}
    
In this formula, $C^{j}$ represents the image corresponding to channel $j$ of a MCMIPF product, $j \in [1,16]$, so $min(\textbf{C})$ is the minimum value of the whole image of channel $j$ and $max(\textbf{C})$ is the maximum value of the whole image of channel $j$, and $\textbf{x}$ represents the $3 \times 3$ image cropped around the point. 

The final product is a table with spatially and temporally co-located data containing two columns: the first one gives the label of the observed cloud type for a given point, and the second contains a matrix of size $3 \times3 \times 16$ containing a GOES-16 image clipping for each point. The realization of this co-location was based on the work of \cite{Gorooh2020}. \rev{This product can then be used}, for example, as an input to an artificial neural network.
    
\begin{figure}[b]
    \centerline{\includegraphics[width=\linewidth]{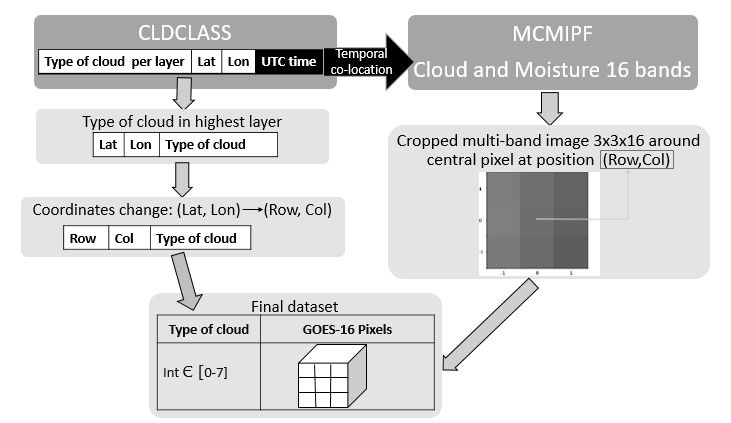}}
    \caption{Schematic diagram of the developed algorithm, which outputs the final product.}
    \label{fig:workflow}
\end{figure}

Fig. \ref{fig:workflow} exhibits a schematic diagram of the process described above to obtain the final product. 
    
\subsection{Verification using a neural network}\label{sec:verify}

To \revv{verify if the co-locations are reasonable}, we used the table as feed to a supervised artificial neural network, \rev{aiming} to generate a thematic map with the cloud types, \revv{and then we compared the map against a True Color product}.
    
The network architecture is linear and contains five layers, with ReLU activation function in between the layers, and a \textit{Softmax} activation function after the output layer. \textit{Cross Entropy Loss} was used as the loss function \cite{crossentropy} and \textit{Adam} \cite{adam} as the optimizer. This architecture is based on the work of Gorooh et al. \cite{Gorooh2020}.
    
We conducted a training in mini-batches of size $50$. The neural network was trained for $250$ epochs at a constant learning rate of $0.001$ using weight decay of $0.00001$, gradient clipping at $0.01$ and Dropout between hidden layers with a probability of $10\%$ as regularization techniques.

\section{Results}
\label{section:results}
        
    The final product has a table structure with two columns. The first column contains, for each row, an integer between 0 and 8 representing the type of cloud that the radar \revv{measures} at a point according to the code summarized in \rev{Table}~\ref{tab:tiposnubes}. The second column contains, for each row, a matrix of size $3 \times 3 \times 16$, the cropped multi-band image around the point for which the cloud type is indicated (see Fig.~\ref{fig:figure_ref}). That is, in each row the first column gives the label of the crop in the second column. The label is co-located with the image crop. This is achieved through the process described in the previous section. 
    
\begin{figure}[t]
    \centerline{\includegraphics[width=\linewidth]{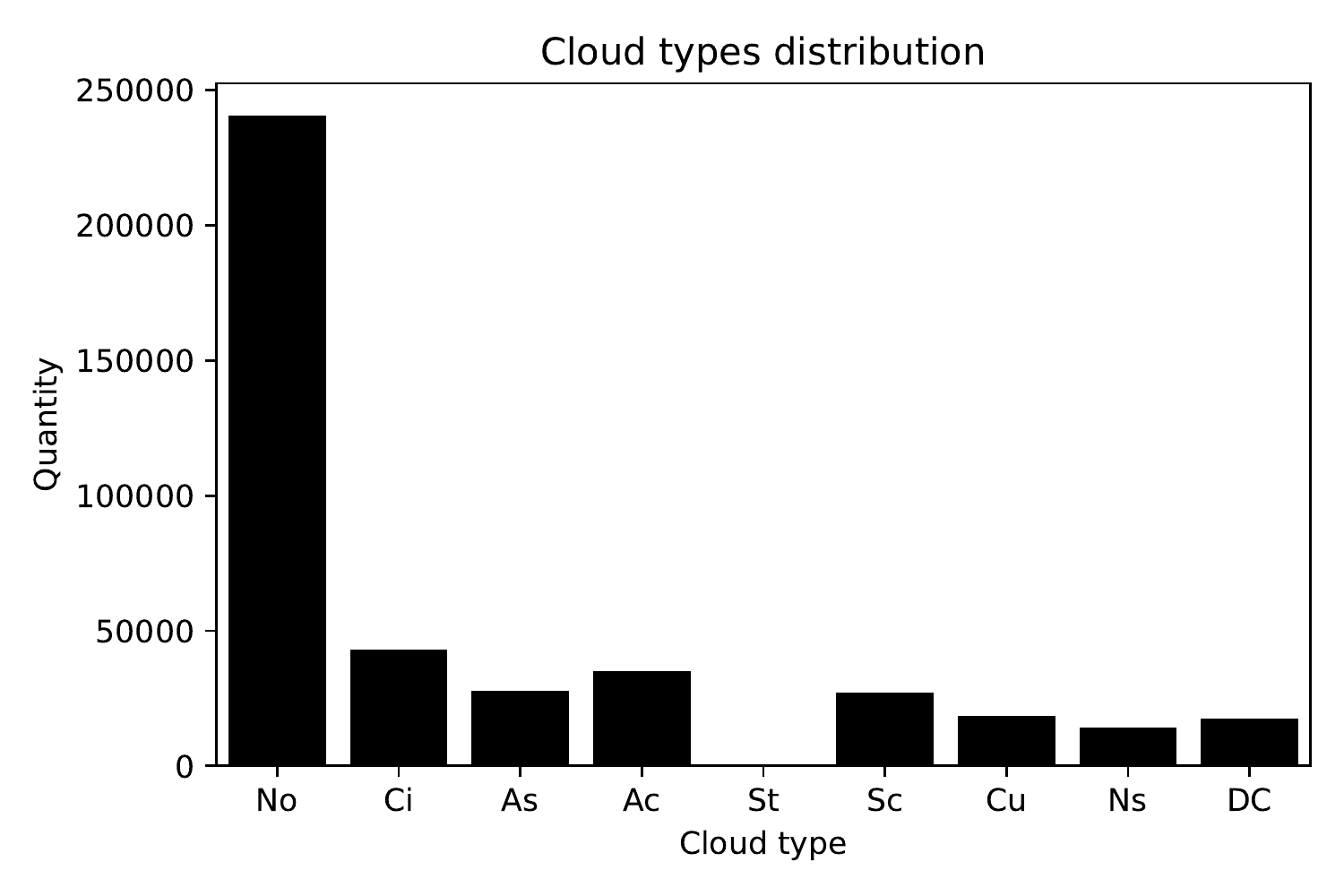}}
    \caption{Distribution of the samples of the data set in the classes available for cloud classification provided by the CLDCLASS product. The x-axis shows the cloud types (see \rev{ Table~\ref{tab:tiposnubes}}) and the y-axis shows the number of samples of each cloud type that exist in the dataset.}
    \label{fig:distribucion}
    \end{figure}

    From the 91 pairs of files, more than 250000 samples were obtained and the classes are in-homogeneously distributed as shown in Fig.~\ref{fig:distribucion}. The most abundant class is the \textit{No-cloud} and the least abundant is the \textit{Stratocumulus} (Sc) class.
    
The algorithm that generates the final product is straightfoward and easy to use, and performs the correct co-locations within the defined spatial and temporal thresholds.

\begin{figure}[t]
\centerline{\includegraphics[width=\linewidth]{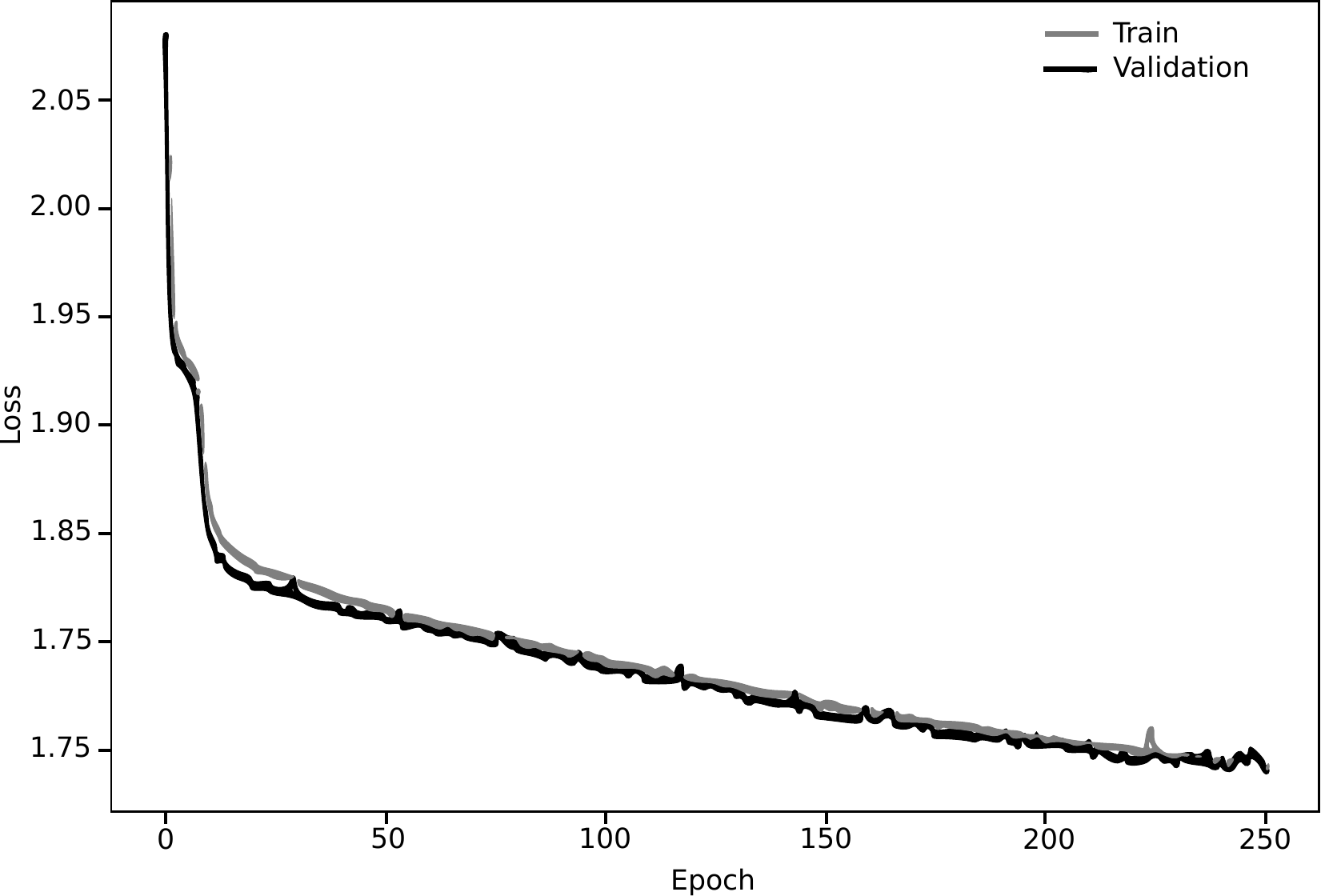}}
\caption{Loss function for every epoch of training. The x-axis shows the epoch of training and the y-axis show the value of the Loss function (or criterion).}
\label{fig:loss}
\end{figure}

Regarding the verification process of the table, with the hyper-parameters defined in section \ref{sec:verify}, an average accuracy of $53\%$ was achieved in both the training and validation sets, and a loss of less than $1\%$ in the last epoch. The classes that obtained the highest precision (This is where the classes assigned by the classifier correspond to the actual class of the cloud) were the ``No clouds'' class and the DC class. The evolution of the loss function over the training epochs for the training and validation sets is shown in Fig.~\ref{fig:loss}. 
    
\begin{figure}[htbp]
\centerline{\includegraphics[width=\linewidth]{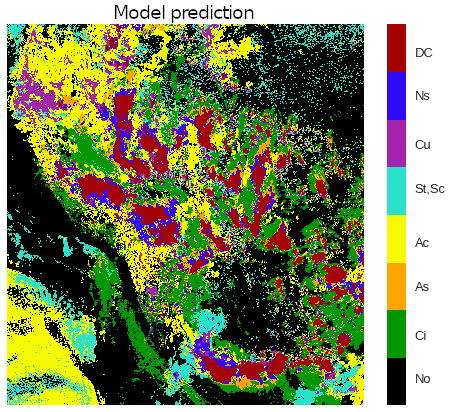}}
\caption{Thematic map with \rev{its} color code, \rev{for a great portion of Southamerica}, produced by the classifier trained by the artificial neural network. Comparing with \rev{Figure}~\ref{fig:realcolor}, predictions of the classifier for deep convection clouds (in red) are, in general, accurate. The classifier overestimates \textit{Cirrus} clouds (in green) in some areas.}
\label{fig:mapatematico}
\end{figure}

\begin{figure}[htbp]
\centerline{\includegraphics[width=\linewidth]{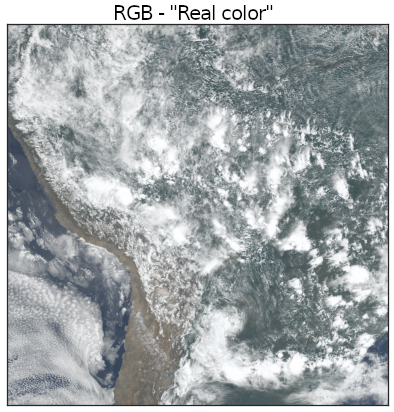}}
\caption{True Color image. In general, clouds are seen in white, the ocean in dark blue and the ground in shades of brown and green. DC clouds are very bright, while \textit{Cirrus} clouds appear dull gray. \textit{Cumulus}-type clouds are observed in the form of ``clusters''. \textit{Stratus} clouds are observed with a smooth texture covering large areas of hundreds of kilometers in extent. Low clouds and fog appear uniformly bright due to the high albedo of the cloud tops compared to the albedos of the ground and water \cite{ilcev} \cite{kidder}.}
\label{fig:realcolor}
\end{figure}

Fig.~\ref{fig:mapatematico} contains the thematic map product of the classifier, while Fig.~\ref{fig:realcolor} contains the True-Color image that attempts to simulate what the human eye would see from the satellite. In the True-Color image the DC clouds are seen in bright white and in the thematic map in red, by visually comparing them we can see that they match between the areas depicted in both images. It can also be seen that the classifier distinguishes quite well the areas with and without clouds. Finally, the \textit{Cirrus} are overestimated in some areas.

\section{Conclusions}
\label{section:conclu}
    
We were able to co-locate data from the ABI radiometer on board a geostationary satellite, GOES-16, and the CPR radar aboard a polar satellite, CloudSat, thus obtaining labels for the multi-band imaging pixels along the each pass over the area of South America captured by the radiometer.
    
Furthermore, we implemented a simple neural network that produced a classifier that successfully distinguishes cloudy pixels from non-cloud pixels and, among the cloudy pixels, successfully distinguishes some cloud types, in particular DC clouds. This implies that the co-location of the data is accurate within some threshold. 
    
\subsection{Further work}
    
The process of selecting the GOES-16 files to download, so that they coincide with the CloudSat pass time over South America, was performed manually. As an improvement, an algorithm could be implemented that would download these files automatically and thus improve the accuracy of the temporal co-location.

On the other hand, the GOES-16 files take up a lot of memory space since the only ABI mode that includes South America is Full Disk, which comprises 16 images of 5424 pixels. However, only a small portion of such an image is used, the 8 pixels around around each central CloudSat pass pixel. One could work on an improvement of the algorithm so that it trims and saves only what is needed from the GOES-16 multi-band image and discards the rest. 

Considering the implementation of these improvements, the temporal window of study could be increased since there is so much data available and the constraints of human work time and memory space would be reduced allowing much more data to be used.


\small
\section*{Acknowledgments}

This work was partially supported by the Consejo Nacional
de Investigaciones Cient\'ificas y T\'ecnicas (CONICET, Argentina), the Facultad de Matem\'atica, Astronom\'ia, F\'isica y Computaci\'on of the Universidad Nacional de C\'ordoba (FaMAF-UNC) and Comisi\'on Nacional de Actividades Espaciales (CONAE, Argentina). P.R.J was supported by a fellowship from CONICET and CONAE, as part of his doctorate at the Instituto Gulich (IG-CONAE-UNC)

This research employed the data from https://www.kaggle.com, Kaggle is an online community for data scientists and machine learning practitioners, as well as the Python programming language, the Numpy, Scipy and GeoPandas libraries.

\normalsize

\bibliographystyle{IEEEtran}
\bibliography{IEEEabrv, mybibfile.bib}

\end{document}